# Femtosecond resolution timing jitter correction on a TW scale Ti:sapphire laser system for FEL pump-probe experiments

**Marta 'Csatari' Divall,[1,*] Patrick Mutter,[1] Edwin J. Divall,[1] Christoph P. Hauri[1,2]**

[1] *Paul Scherrer Institut, 5232 Villigen PSI, Switzerland*
[2] Ecole Polytechnique Fédérale de Lausanne, 1015 Lausanne, Switzerland
[*] marta.divall@psi.ch

**Abstract:** Intense ultrashort pulse lasers are used for fs resolution pump-probe experiments more and more at large scale facilities, such as free electron lasers (FEL). Measurement of the arrival time of the laser pulses and stabilization to the machine or other sub-systems on the target, is crucial for high time-resolution measurements. In this work we report on a single shot, spectrally resolved, non-collinear cross-correlator with sub-fs resolution. With a feedback applied we keep the output of the TW class Ti:sapphire amplifier chain in time with the seed oscillator to ~3 fs RMS level for several hours. This is well below the typical pulse duration used at FELs and supports fs resolution pump-probe experiments. Short term jitter and long term timing drift measurements are presented. Applicability to other wavelengths and integration into the timing infrastructure of the FEL are also covered to show the full potential of the device.

## References and links

1. S. Witte and Kjeld S. E. Eikema, "Ultrafast Optical Parametric Chirped-Pulse Amplification," IEEE JQE vol.18, Issue 1, pp. 296 – 307 (2012)
2. A. Schwarz, M. Ueffing, Y. Deng, X. Gu, H. Fattahi, T. Metzger, M. Ossiander, F. Krausz, and R. Kienberger, "Active stabilization for optically synchronized optical parametric chirped pulse amplification," Opt. Express 20, 5557-5565 (2012)
3. H. Zeng, H. Xu, K. Wu, E. Wu, "Generation of accurately synchronized pump source for optical parametric chirped pulse amplification," Appl. Phys. B 79, 837-839 (2004)
4. Major, Z., Trushin, S., Ahmad, I., Siebold, M., Wandt, C., Klingebiel, S., et al. "Basic concepts and current status of the petawatt field synthesizer - a new approach to ultrahigh field generation," Laser Rev., vol. 37, 431–436 (2009)
5. F. Batysta, R. Antipenkov, J. T. Green, J. A. Naylon, J. Novák, T. Mazanec, P. Hříbek, C. Zervos, P. Bakule, Be. Rus, "Pulse synchronization system for picosecond pulse-pumped OPCPA with femtosecond-level relative timing jitter," Opt. Express 22, 30281-30286 (2014)
6. S. Prinz, M. Häfner, M. Schultze, C. Y. Teisset, R. Bessing, K. Michel, R. Kienberger, T. Metzger, "Active pump-seed-pulse synchronization for OPCPA with sub-2-fs residual timing jitter," Opt. Express **22**, 25, 31050-31056 (2014)
7. T. Miura, F. Kannari, K. Takasago, and K. Torizuka, " Timing stabilized regenerative amplifier with spectral-resolved cross-correlation technique," in *Advanced Solid-State Lasers*, C. Marshall, ed., Vol. 50 of OSA Trends in Optics and Photonics (Optical Society of America, 2001), paper WB3.
8. J. M. Glownia, J. Cryan, J. Andreasson, A. Belkacem, N. Berrah, C. I. Blaga, C. Bostedt, J. Bozek, L. F. DiMauro, L. Fang, J. Frisch, O. Gessner, M. Gühr, J. Hajdu, M. P. Hertlein, M. Hoener, G. Huang, O. Kornilov, J. P. Marangos, A. M. March, B. K. McFarland, H. Merdji, V. S. Petrovic, C. Raman, D. Ray, D. A. Reis, M. Trigo, J. L. White, W. White, R. Wilcox, L. Young, R. N. Coffee, and P. H. Bucksbaum, "Time-resolved pump-probe experiments at the LCLS," Opt. Express 18, 17620-17630 (2010)
9. M. P. Minitti, J. S. Robinson, R. N. Coffee, S. Edstrom, S. Gilevich, J. M. Glownia, E. Granados, P. Hering, M. C. Hoffmann, A. Miahnahri, D. Milathianaki, W. Polzin, D. Ratner, F. Tavella, S. Vetter, M. Welch, W. E. White and A. R. Fry*," Optical laser systems at the Linac Coherent Light Source," J. Synchrotron Rad. 22, 526-531 (2015)

10. S. Schulz, L.-G. Wissmann, V. Arsov, M.K. Bock, M. Felber, P. Gessler, K. E.Hacker, F. Ludwig,H. Schlarb, B. Schmidt, J. Zemella, "Precision synchronization of the Flash Photoinjector Laser," WEPWB076 IPAC10, Kyoto (2010)
11. T. J. Maxwell, P. Piot, J. Ruan, M. J. Kucera, "Synchronization and jitter studies of the Titanium-Sapphire laser at the A0 photoinjector," MOP285 Proc PAC11, New York (2011)
12. J. Hong, J.-H. Han, S.J. Park, Y.G. Jung, D.E. Kim, H.-S. Kang, and J. Pflueger," A study on low emittance injector and undulator for PAL-XFEL," High Power Laser Science and Engineering, **3**, e21 (2015)
13. M. B. Danailov, M. B. Alsous, P. Cinquegrana, A. Demidovich, G. Kurdi, I. Nikolov, P. Sigalotti, "Study of a collinear single-shot-type cross-correlator for laser timing applications, " Appl. Phys. B **120**, 1, pp 97-104 (2015)
14. A. Trisorio, P. M. Paul, F. Ple, C. Ruchert, C. Vicario, C. P. Hauri, "Ultrabroadband TW-class Ti:sapphire laser system with adjustable central wavelength, bandwidth and multi-color operation," Opt. Express **19**, 20128-20140 (2011)
15. C.P. Hauri, R. Ganter, B. Beutner, H.H. Braun, C. Gough, R. Ischebeck, F. Le Pimpec, M.L., Paraliev, M. Pedrozzi, C. Ruchert, T. Schietinger, B. Steffen, A. Trisorio, C. Vicario, "Wavelength-tuneable UV laser for electron beam generation with low intrinsic emittance, " Proc. IPAC10 WEPD052, Kyoto, Japan (2010)
16. M. Csatari Divall, M. Kaiser, S. Hunziker, C. Vicario, B. Beutner, T. Schietinger, M. Lüthi, M. Pedrozzi, and C. P. Hauri, "Timing jitter studies of the SwissFEL test injector drive laser," Nucl. Instrum. Methods Phys. Res., Sect. A **735**, 471 (2014)
17. C. J. Milne (ed.) "Experimental Station A: Conceptual Design Report SwissFEL, " http://www.psi.ch/swissfel/
18. G. Ingold and P. Beaud (ed.), "Conceptual Design Report SwissFEL ARAMIS Endstation ES-B, " http://www.psi.ch/swissfel/CurrentSwissFELPublicationsEN/ES-B_CDR_2013-06-21_v2_with_coverpage_%282%29-VM16.pdf
19. A. L. Cavalieri et al. " Clocking femtosecond X-rays, " Phys. Rev. Lett. **94**, 114801 (2005).
20. P.N. Juranić, A. Stepanov, P. Peier, C.P. Hauri, R. Ischebeck, V. Schlott, M. Radovi , C. Erny, F. Ardana-Lamas, B. Monoszlai, I. Gorgisyan, L. Patthey, R. Abela, " A scheme for a shot-to-shot, femtosecond-resolved pulse length and arrival time measurement of free electron laser x-ray pulses that overcomes the time jitter problem between the FEL and the laser, " JINST 9 P03006 (2014)
21. A. Winter, E-A. Knabbe, S. Simrock, B. Steffen, N. Ignashin, A. Simonov, S. Sytov, "Femtosecond synchronisation of ultrashort pulse lasers to a microwave RF clock, " RPAT094 Proc. of PAC05, Knoxville, Tennessee (2005)
22. V. Arsov, S. Hunziker, M. Kaiser, V. Schlott, F.Loehl, "Optical synchronization of the SwissFEL 250MeV test injector gun laser with the optical master oscillator, " TUPA21 Proc FEL2011, Shanghai, China (2011)
23. M. Xin, K. Şafak, M. Y. Peng, P. T. Callahan, F. X. Kärtner, "One-femtosecond, long-term stable remote laser synchronization over a 3.5-km fiber link," Opt. Express **22**, 14904-14912 (2014)
24. S. Hunziker, V. Arsov, F. Buechi, M. Kaiser, A. Romann, V. Schlott, P. Orel, S. Zorzut "Reference distribution and synchronization systems for SwissFEL: Concept and First results," IBIC, MOCZB, Monterey, Ca., USA, (2014)
25. S. Schulz, I. Grguras, C. Behrens, H. Bromberger, J.T. Costello, M. K. Czwalinna, M. Felber, M. C. Hoffmann, M. Ilchen, H.Y. Liu, T. Mazza, M. Meyer, S. Pfeiffer, P. Predki, S. Schefer, C. Schmidt, U. Wegner, H. Schlarb, A. L. Cavalieri, "Femtosecond all-optical synchronization of an X-ray free-electron laser, " Nat. Comm., **6**, 5938 (2015)
26. P. Cinquegrana, S. Cleva, A. Demidovich, G. Gaio, R. Ivanov, G. Kurdi, I. Nikolov, P. Sigalotti, and M. B. Danailov, "Optical beam transport to remote location for low jitter pump-probe experiments with Free Electron Laser," Phys. Rev. ST Accel. Beams 17(4), 040702 (2014).
27. P. E. Ciddor, "Refractive index of air: new equations for the visible and near infrared," Appl. Opt. Vol. 35, pp. 1566–1573 (1996)
28. V. Arsov, M. Aiba, M. Dehler, F. Frei, S. Hunziker, M. Kaiser, A. Romann,V. Schlott "Commissioning and results from the bunch arrival-time monitor downstream the bunch compressor at the SwissFEL injector test facility, " Proceedings of FEL2014, Basel, Switzerland THP085
29. B. Schmidt, M. Hacker, G. Stobrawa, T. Feurer; LAB2-A virtual femtosecond laser lab, http://www.lab2.de

---

## 1. Introduction

High power ultrashort pulse lasers are used in many area of research, where high time-resolution is required. Ti:sapphire based chirped pulse amplifiers (CPA) provide fs pulses and a well established, robust technology at 10's of mJ output energy. Newly rising short pulse pumped OPCPA systems extend the range of energy per pulse available. As pointed out by Witte et al. [1] the synchronization between the seed laser, - typically Ti:sapphire laser,- and the pump laser to sub-ps accuracy is crucial for these systems. This has prompted the development of time of arrival measurement and correction systems for the entire laser chain [2,3,4,5,6]. The spectrally resolved cross-correlation solution [7] applied in these publications looks especially attractive and promising, when broadband pulses are available.

High power Ti:sapphire lasers are also frequently used as part of large scale machines, such as free electron lasers (FEL) and accelerators, where they are used for the generation of secondary sources, diagnostics and for pump-probe experiments [8-13]. It is also the chosen technology for SwissFEL [14-18]. To study ultrafast dynamics of material structures, with excitation lifetimes on the fs scale requires the time of arrival of the laser pulses and the X-rays to be perfectly timed on the fs scale. The optical pump laser has been shown to jitter in the 100 fs region and is the main limitation to the real time time-resolution of these experiments [8]. Binning techniques have been developed to post-process data [19, 20]. Here the time-resolution depends on the time-window provided by the binning tool. Therefore if the time of arrival of optical laser is actively stabilized to a higher level, this provides better binning resolution, higher rate of useful data and better online analysis options. Therefore the aim is to improve the overall timing stability of the laser system.

Optical synchronization of the seed oscillator of the laser chain to the machine's optical reference has been improved over the years and is now demonstrated down to the 10fs level, using balanced optical cross-correlation (BOC) [20,23]. This provides a straightforward timing in comparison to the machine frequencies distributed with high stability over fiber-links [23, 24, 25]. However, the rest of the laser chain typically consists of multi-stage amplification, regenerative amplifier cavities, active elements as well as a long, 10's of meters propagation to the FEL experiment which makes it prone to long term timing drift and fast jitter sources which in turn leads to several 100 fs error in the timing of the delivered pulse on target [16, 26].

Our aim was to adapt the existing cross-correlation based time-or-arrival measurement for a high power laser system used in SwissFEL, to reach sub-10fs resolution and provide an active control to keep the laser output in time with the optical clock, serving as the heart of the machine. Therefore we developed a single shot, optical non-collinear cross-correlator with fs resolution, operating at 5Hz, upgradable to 100Hz, which is the repetition rate of SwissFEL. The device keeps the output of our 100Hz TW class Ti:sapphire amplifier chain [14-16] in time with the seed oscillator to below 3 fs RMS level over several hours. Shot to shot, short term jitter and long term timing drift measurements are presented, with discussion on the sources of the noise. The device is currently used at 800nm. The adaptation of the system to compare against the 1550 nm reference laser is also discussed. This will enable drift compensated, fs resolution pump-probe experiments at SwissFEL.

## 2. The cross-correlator design

The time of arrival tool has to fulfil several requirements. The device has to have a resolution below 10 fs and be at least an order of magnitude more stable by itself over time. It has to provide a feedback signal to an active delay in the system, to keep the timing on the target constant. It has to be remotely adjustable where direct access during operation is difficult. Finally, it has to be integrated into the overall timing system of the FEL, as well as the associated control system, in our case Experimental Physics and Industrial Control System (EPICS). The following sections summarize the strategies followed to reach these targets.

### *2.1 Optical setup of the single shot cross-correlator*

The optical setup is shown on Fig.1. It has been designed to allow for spectrally resolved cross-correlation measurements (a). One arm receives the optical reference pulse, which in the future will be the distributed optical clock (OneFive, 1550nm), currently the seed oscillator (Femtolaser, 800nm) of the laser system. The reference pulse is then stretched by a passive optical element (eg. $CaF_2$ block) to apply a well-defined chirp. The second arm receives the pulses to be stabilized. A small fraction of the compressed output from the laser was used in our case. The two arms are then non-collinearly mixed on a nonlinear crystal to generate a sum-frequency signal (type I BBO SHG at 800nm $\Theta = 29.2°$;$\Phi = 0°$ in the applied case). This beam is then spatially and spectrally filtered from the initial beams and their generated second-harmonics, ready for observation. The external angle of the two beams is set to 3°. The crystal can be rotated and translated remotely, to find the tuning angle and the spatial overlap.

One can either measure the intensity of the mixed signal, which will give cross-correlation, or the spectrum of the signal, which will give the relative timing of the signals, independent of the intensity fluctuation. In this case the relative timing translates to wavelength change of the mixed signal, which will appear as a spatial displacement on the spectrometer. For the latter, the beam was collimated after the crystal with a 10cm focal-length lens. The first order reflection from a UV grating with 2400gr/mm was imaged onto the camera using a 40mm focal-length lens. The setup can also be applied as an auto-correlator to measure the dispersion applied. The setup was modified for the feedback stabilization (b). Due to the low power available from the oscillator (5mW) the grating was replaced by a mirror and simple cross-correlation was performed.

For observation of the mixed signals we have used a CCD camera (Basler, scA640-120gm, 5.6µm square pixels), using in house software we can simultaneously observe the peak positon of the spectrum and the integration of the whole signal, using region of interest (ROI), see Fig.2. The acquisition rate of the camera is 100 Hz, which is sufficient for our 100Hz operational frequency. For the presented measurements the network data-rate limited us to 5 Hz, but moving the analysis to run on the same PC as the capture this can be increased to 100Hz in the future. The camera is also useful for remote setup of the device, as it allows visualization of the beams from both arms and selection of the correct tuning angle. To ensure temporal overlap of the signals a linear motor-stage is used (SmarAct SLC-24150). The beam height is kept the lowest possible (7.5cm) to avoid vibrational instability. The total footprint is $60X60cm^2$.

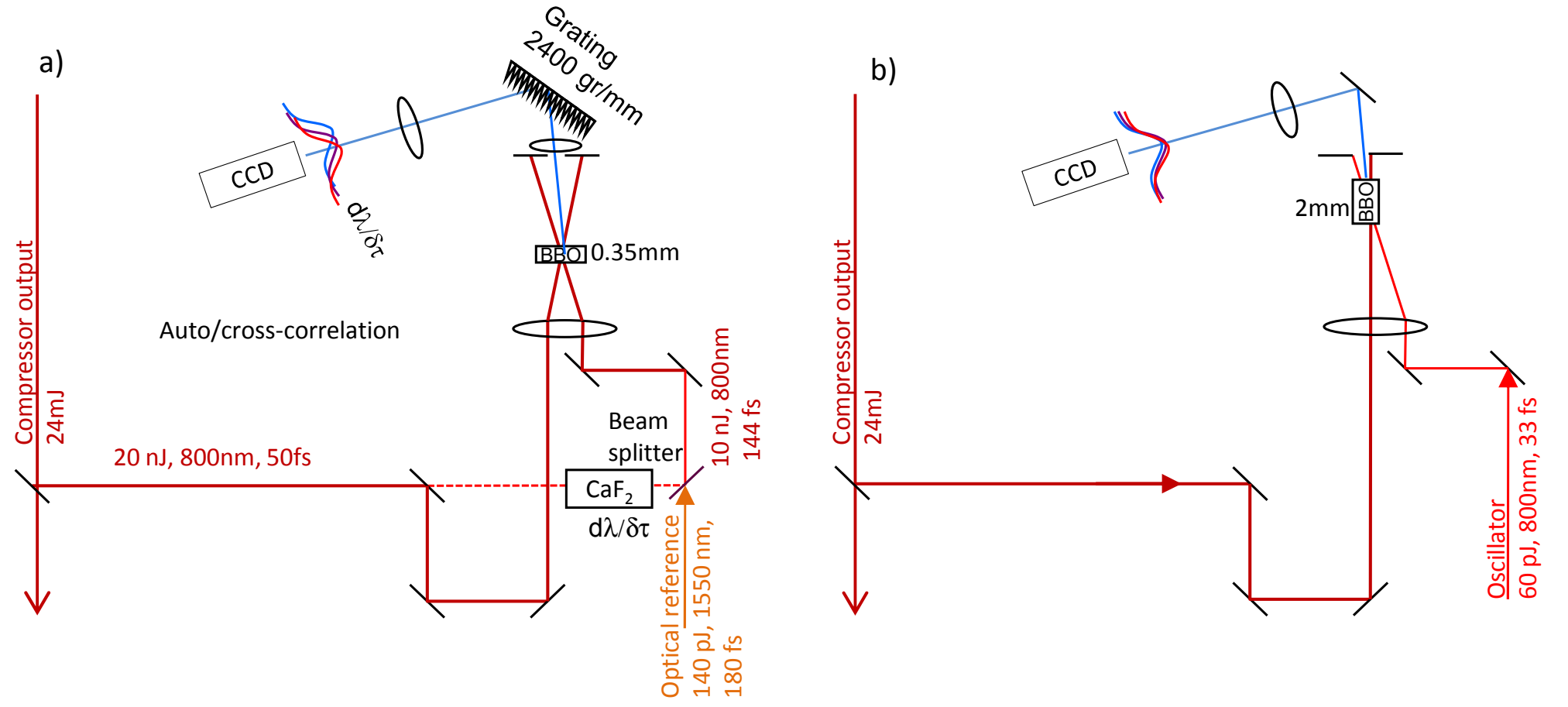

Fig. 1. Setup of the optical cross-correlator. (a) spectrally resolved cross-correlation; (b) simple cross-correlation

## *2.2 The interface*

The device is operated via EPICS. In Fig.2 the control panel of the LAM is shown. The spectrum is imaged onto the camera and the software the camera parameters, region of interest to be selected and images captured. An IOC based scanning tool takes the calibration curves and records up to 5 data channels for long term stability measurements.

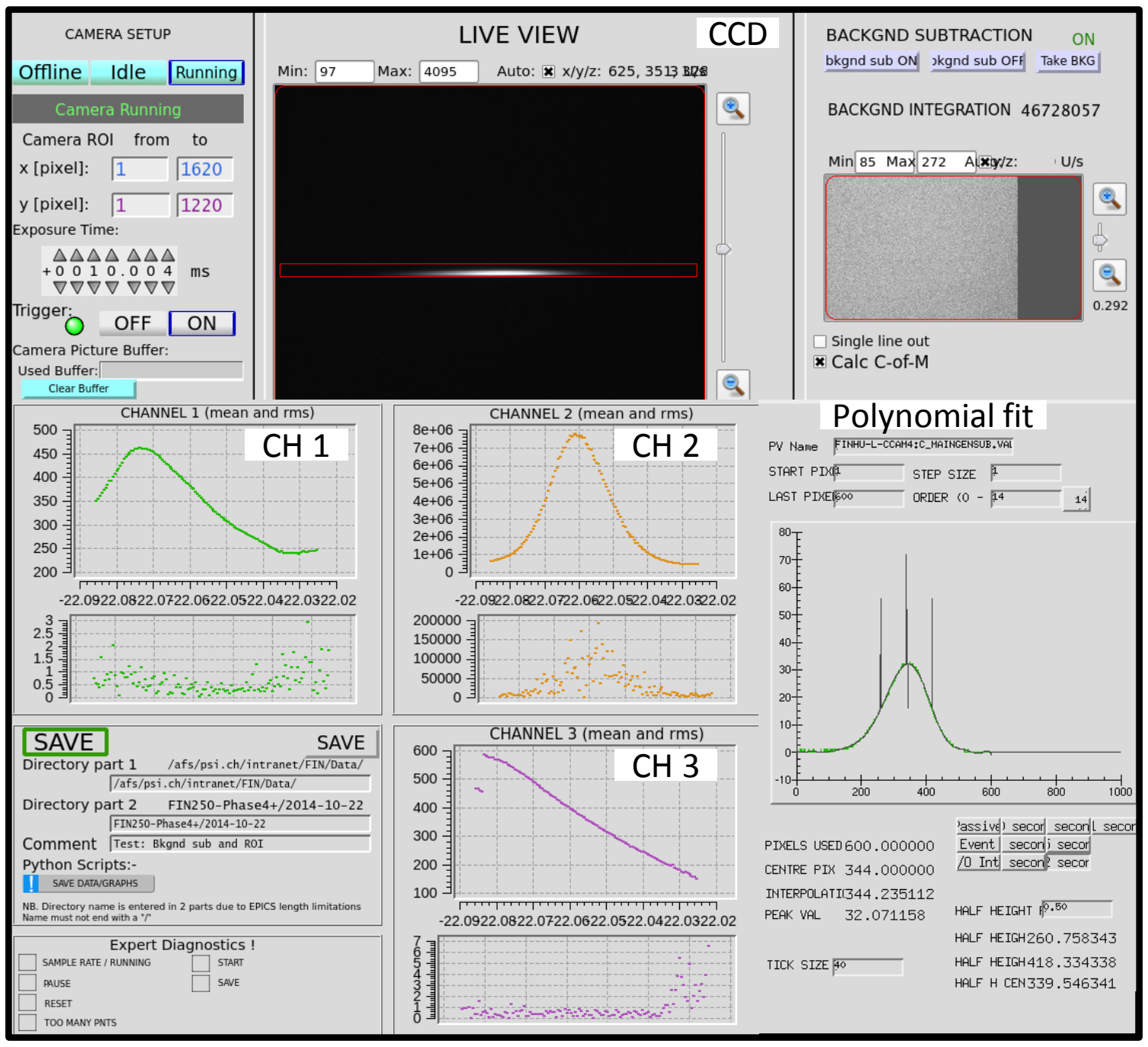

Fig. 2. EPICS interfaces. Top shows the camera viewer and the background subtraction. Bottom left shows the scanning tool, while bottom right the polynomial fit to the spectrum.

On channel one the centroid pixel on the camera corresponding to the centroid wavelength of the mixed spectrum versus the motor position is displayed. In channel two the intensity (integral over the selected region of interest) versus the motor position is recorded, corresponding to the intensity auto/cross correlation. For every sample point we also perform a polynomial fit of the spectral intensity distribution on the camera. The pixel number of the peak intensity is recorded in channel three versus the motor position. This gives a larger linear range for the measurement without compromised resolution, as the peak position of the spectrum can be detected even when the spectrum is at the edge of the camera.

## 3. Timing measurements

Our aim was to determine the resolution of the laser arrival monitor, measure its long term stability, to apply it to our TW laser system for timing jitter measurements and to generate an error signal for feedback stabilization. The best resolution (<1fs) was achieved with the spectrally resolved cross-correlation, with extended range (>200fs) by applying polynomial fit to the spectrum. The modified setup was used for the feedback stabilization of the timing. Results are presented below.

### *3.1 Resolution and stability of the measurement devices*

The available oscillator power did not allow for spectrally resolved measurements, using this signal as a reference. Therefore, the output of the compressor was measured against itself with the spectrally resolved cross-correlator to obtain the resolution of the system in this configuration. The compressed pulse (FWHM = 50 fs) was split in between two arms of the setup and one arm was stretched by a 5cm $CaF_2$ block to 144 fs FWHM. For the mixing a 0.35 mm long type I BBO crystal was used, cut for SHG at 800 nm ($\Theta$=29.2$^o$), anti-reflection coated for 800 nm on the front surface and 400 nm on the back surface.

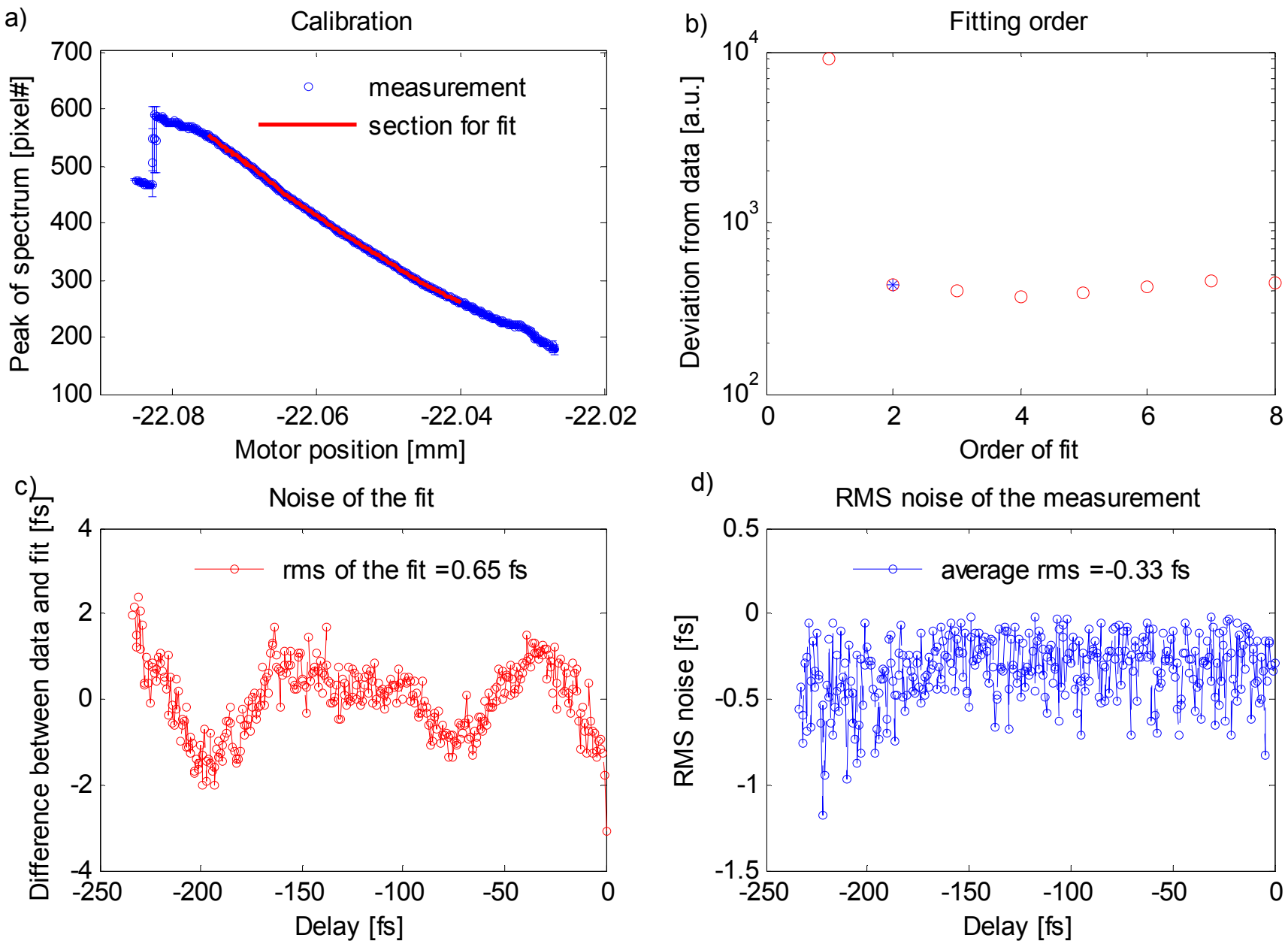

Fig. 3. Calibration curve of the spectrally resolved cross-correlation for the compressor output; (a) calibration curve with the region to be used in red; (b) Error of fit vs polynomial fit order; (c) deviation of the data from the fitting; (d) noise of the data over the range

The top left figure in Fig.3 shows the recorded calibration curve, corresponding to the peak of the polynomial fit to the spectrum versus the delay of the motor. The red curve is the region, where the device is intended to be used. In the top right figure the deviation of the data versus the polynomial fitting order is displayed and the fitting order used for the further analysis is highlighted in blue (parabolic in this case). The two remaining figures quantify the resolution. This scheme gives about 1fs resolution for a measurement range of 230 fs. If the system is used at the central 50 fs range then 0.3 fs resolution is achievable. As we show later it is possible to keep the overlap in this range with the feedback stabilization running.

Long term stability of the setup itself was measured at a fixed delay at the middle of the linear range over 5 hours and is shown in Figure 4. The main contribution to the drift is related to temperature changes in the laser lab. The correlation slope shows a 16 fs/$^{o}$C drift. In SwissFEL the room will be stabilized to 0.1 $^{o}$C accuracy and the temperature of the LAM box will be monitored with a few mK precision. Therefore the system should provide an excellent stability and resolution over time.

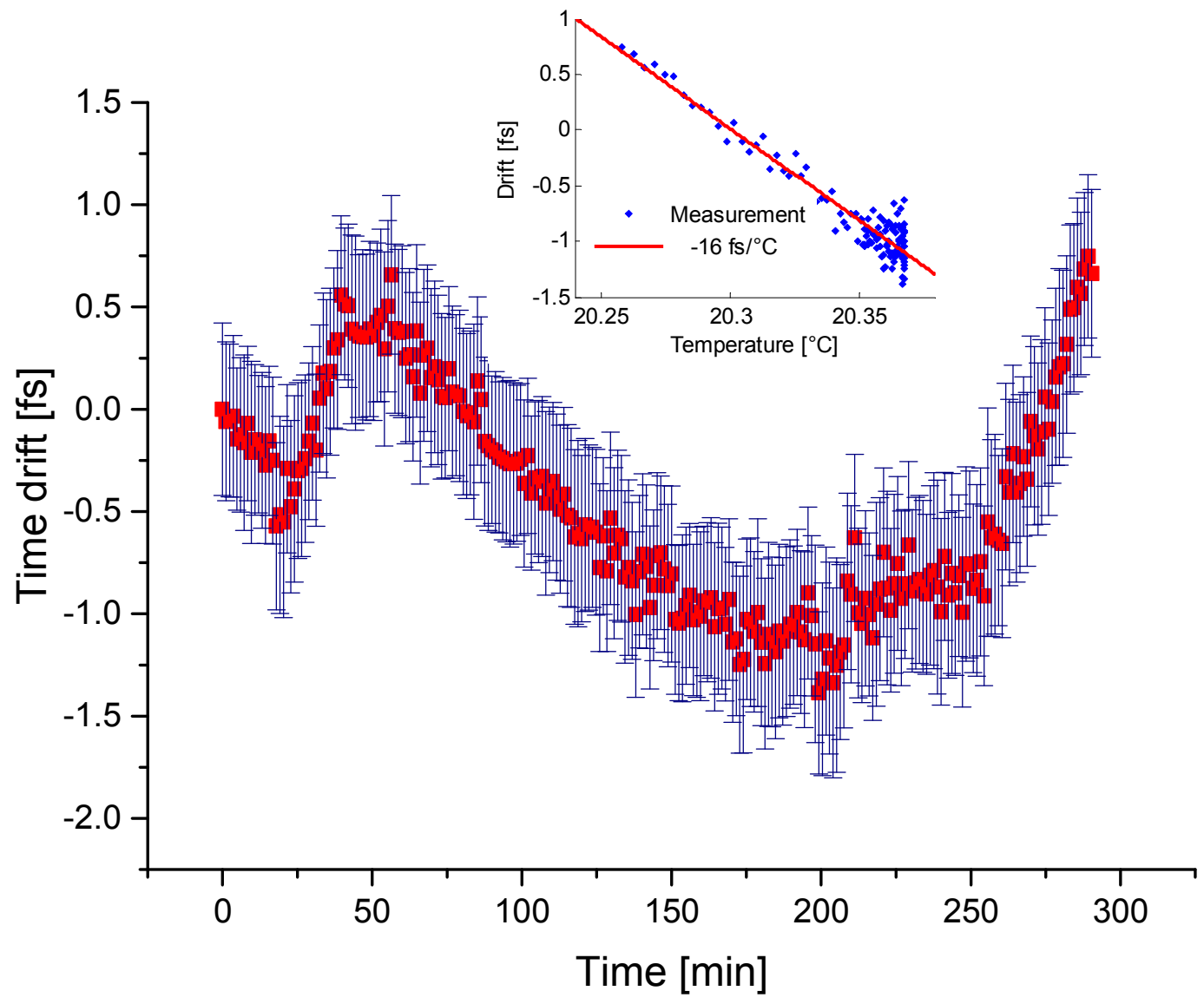

Fig. 4. Long term stability of the spectrally resolved cross-correlator. Drift over 5 hours and the correlation with the room temperature (inset), showing -16fs drift over 1 Celsius change in the room temperature.

For the feedback stabilization the modified setup was used (Fig.1 (b)). Here we compared the compressor output to the nearest coinciding oscillator pulse, which was transported to the output of the compressor over a 5m stable, enclosed transfer line. The laser oscillator was electronically synchronized to an external reference with an excellent ~30fs integrated jitter between 10Hz and 10MHz (Femtolock) [16]. The output of the compressor after a total propagation path in the laser system, including booster amplifier, stretcher, regenerative amplifier and two multi-pass amplifiers, corresponds to ~60m.

The crystal was replaced by a 2mm long BBO and the amplifier beam was collimated through the crystal to increase the overlapping area between the two beams and therefore the interaction length and the signal strength. Spectrally resolved measurements were not possible due to the low signal level from the oscillator, therefore the grating was replaced by an aluminum mirror. The signal was projected onto the camera and the integral of the beam was measured over the selected region of interest. Figure 5. shows the calibration curve of the cross-correlation. The resolution and noise of the system is ~5 fs in the used range, which is larger than for the spectrally resolved configuration, but still allows for long term drift measurements and correction. The full range covers 56 fs. Amplitude variations of the compressor output were recorded simultaneously with LaserProbe Rjp465 energy meter to decouple it from the measurements. The amplitude variation measured would cause less than 1 fs error in the timing and was neglected in this case.

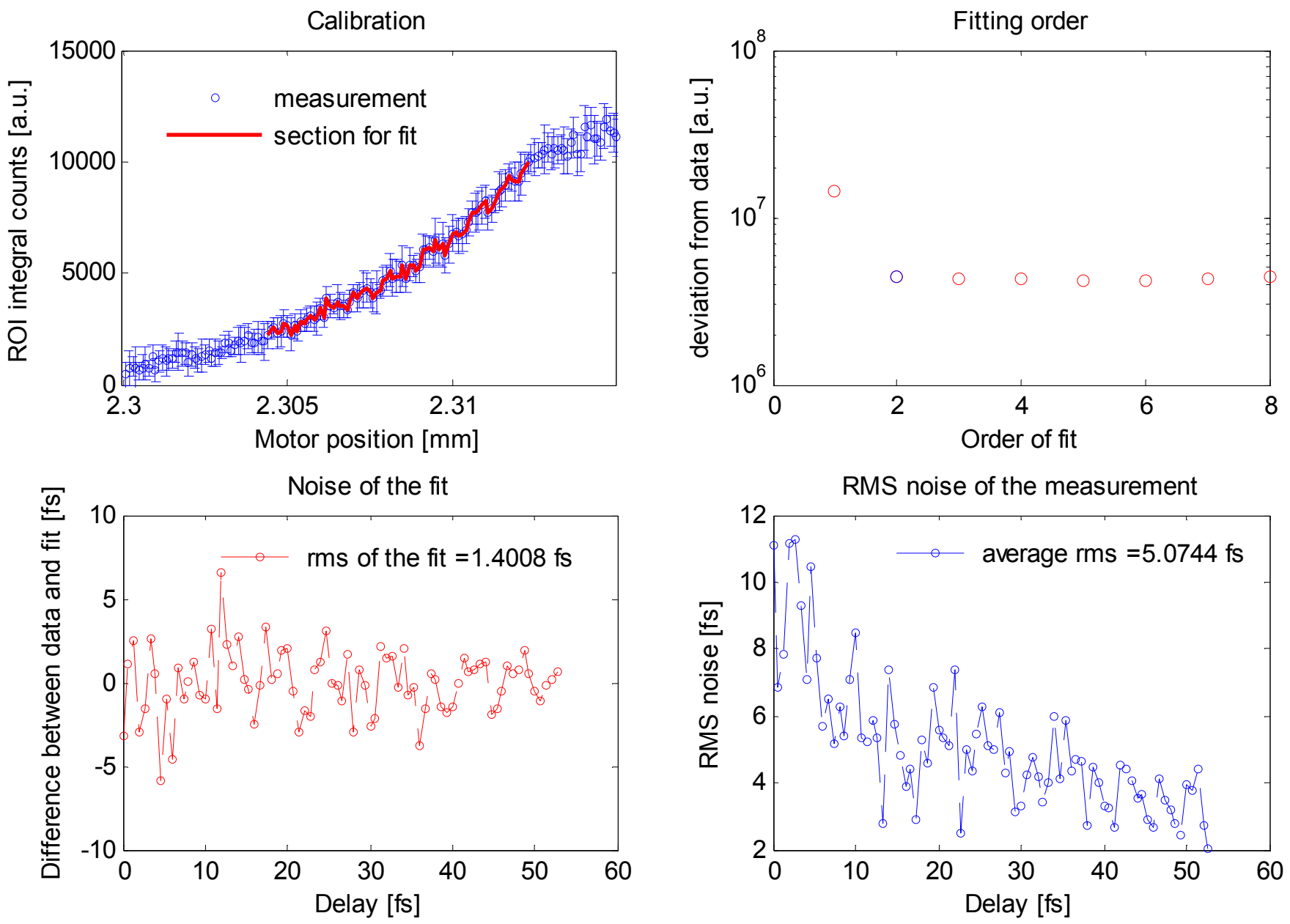

Fig. 5. Calibration curve of the cross-correlation between the compressor output and the seed oscillator; (a) calibration curve with the region to be used in red; (b) error of fit vs polynomial fit order; (c) deviation of the data from the fitting; (d) noise of the data over the range

### *3.2 Timing correction*

An on/off feedback loop was implemented which drives the delay stage after the compressor to keep the pulse overlapped with the oscillator. Polynomial fit parameters of the calibration, tolerance levels for actuation and the number of shots to average over can be selected. Fig 6 shows the feedback setting for the long term measurements. The tolerance bound was set to +/- 4 fs, corresponding to +/-564 counts. If the signal averaged over 5 shots drifts out of these bounds, the motor is moved by a constant step-size of 6 μm. The range of the measurement device (56fs) is set as the upper and lower limit for the loop to turn off.

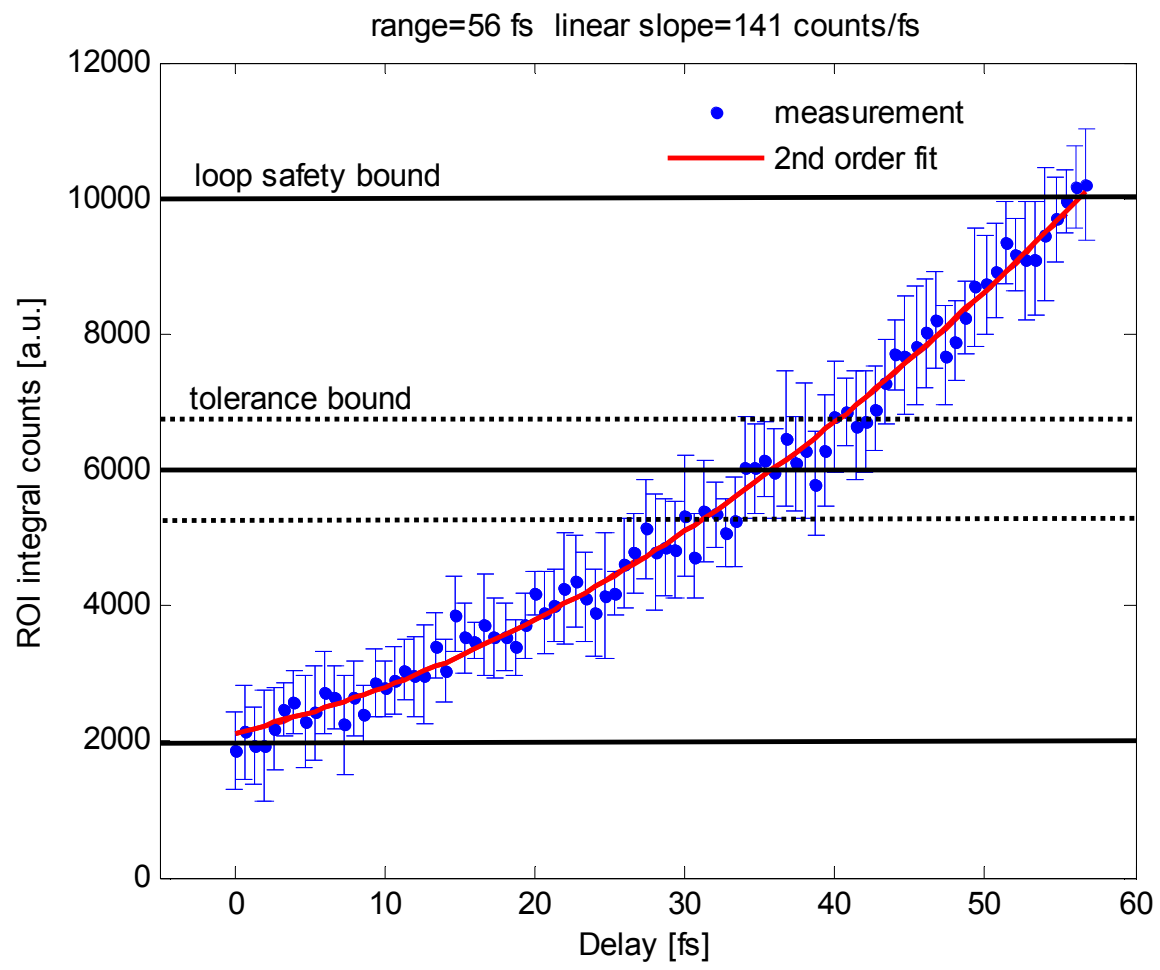

Fig. 6. Calibration curve of the cross-correlator with the tolerance bounds and the working range of the loop.

### *3.3 Timing measurements of the laser system*

Using this simple feedback the overall rms jitter from the oscillator to the compressor output was kept below 3 fs, while the peak to peak deviation was below 10 fs (Fig.7.). The total drift without correction would have been ~300fs over 1 hour and more than 700 fs over the whole 10 hour measurement period. The feedback could have run longer if we did not terminate the measurement.

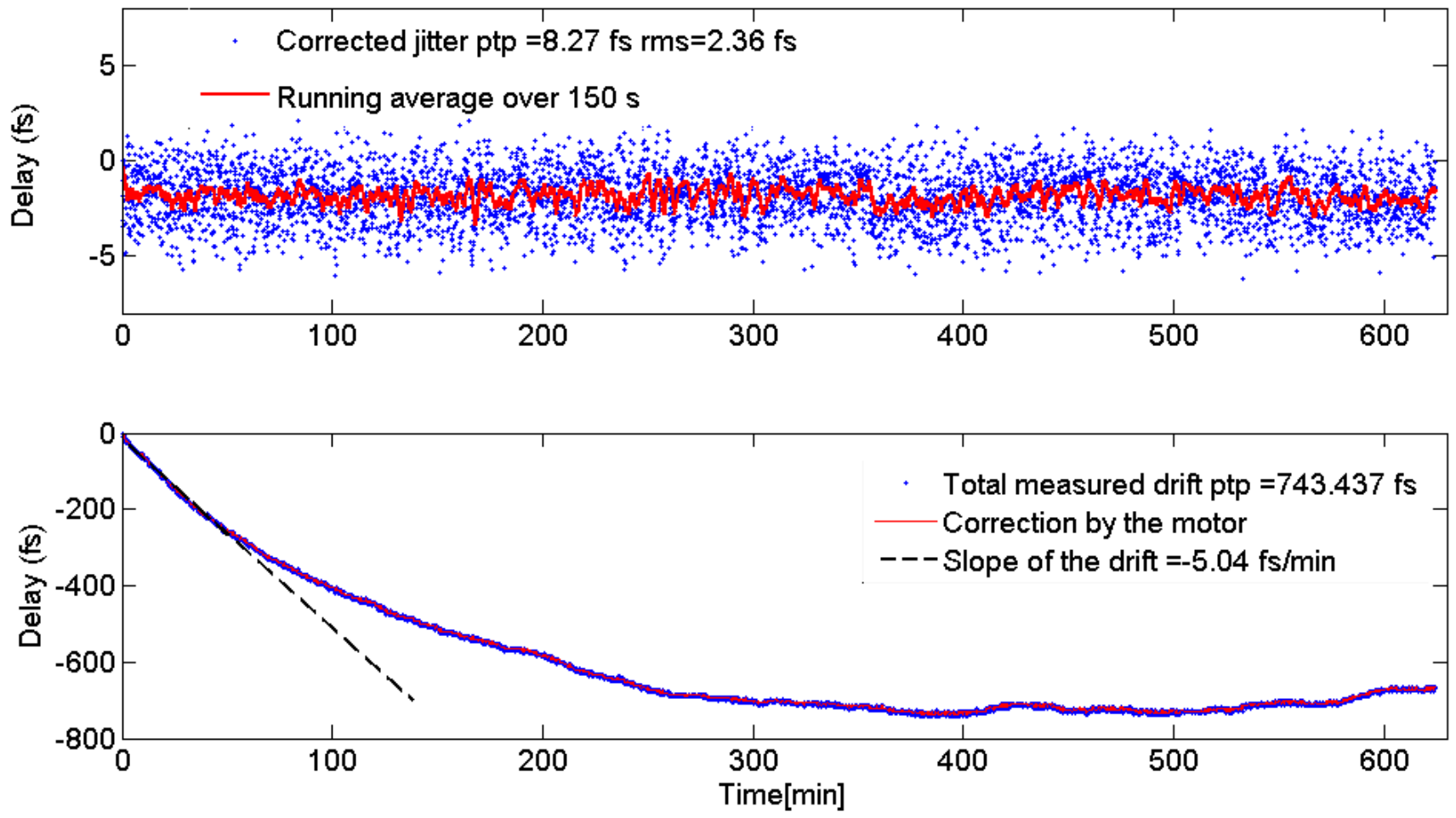

Fig. 6. Long term stability measurement over 10 hours. The corrected timing jitter of the system (top), the drift derived from the correcting motor movement (bottom)

The laboratory conditions were recorded during the measurement, showing a similar trend as the overall drift (Fig.7.), but only accounting for ~ 200 fs change in the delay in air. The rest of the drift comes from thermal expansion of the components in the laser system.

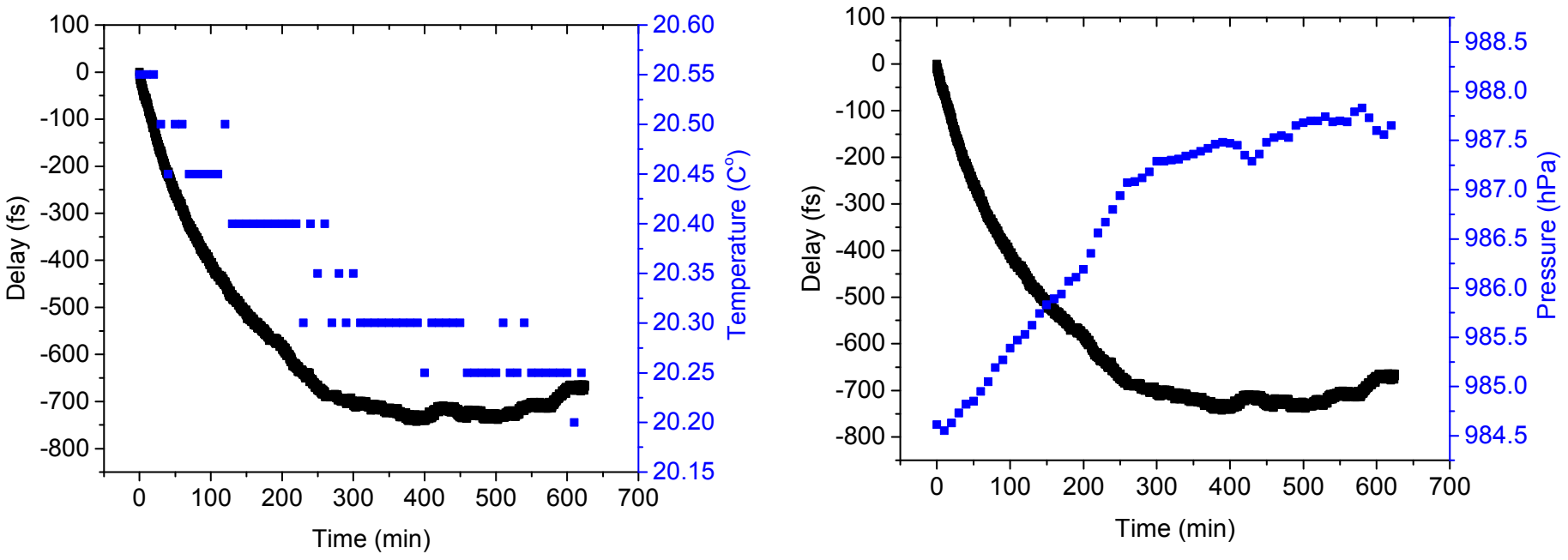

Fig. 7. Correlation with environmental changes. Pressure change during the measurement (left); temperature of the laboratory (right)

An interesting segment of the measurement between 5-6 hours is shown on Fig.8. The peak to peak jitter over this 1 hour period was ~7 fs, while the rms was ~2.5 fs. The periodic variations on a few minutes scale were observed and correlated to the humidity control of the air-conditioning system, giving 18.2fs/%RH slope. This corresponds well to the value predicted from the Ciddor equation [27], giving 0.3fs/%RH/m path-length change due to variation of humidity.

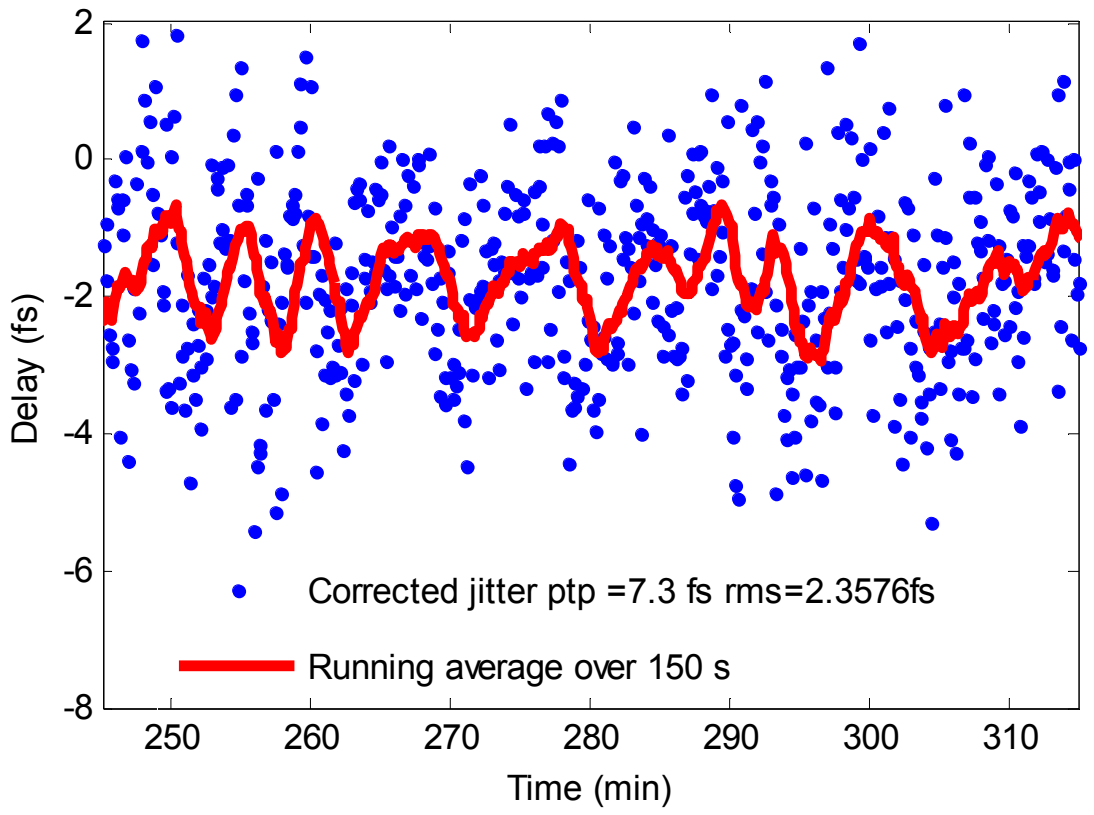

Fig. 8. Short term jitter of laser timing, measured at 5 Hz, with running average over 150s, highlighting the periodic variations related the humidity change in the lab.

## 4. Adaptation and integration into the FEL

The spectrally resolved cross-correlation setup can be adapted to compare the timing of the reference master clock, operating at 1550nm, delivering 180 fs transform limited pulses. 10-20mW power will be available for laser arrival time measurements. To calculate the resolution and the crystal needed for the measurement, we have used LabII [29]. A BBO crystal cut at 22.2$^{o}$ provides the sum-frequency signal at 528 nm. Although difference-frequency provides a higher slope the produced 1650nm signal can be hard to spectrally filter from the optical reference and detection is easier in the visible range. The compressed pulses produced near the FEL experiment will be used as the broadband stretched pulse. The example shown on figure 9 provides a range of +/- 100 fs measurable with a pulse stretched from 30 fs to ~80fs in 2cm of BK7. This temporal range requires 12 nm spectral width to be displayed over 659 pixels of the CCD. With the 0.012nm/fs shift calculated we get 1.5fs/pixel resolution. If the range can be further reduced the resolution would improve.

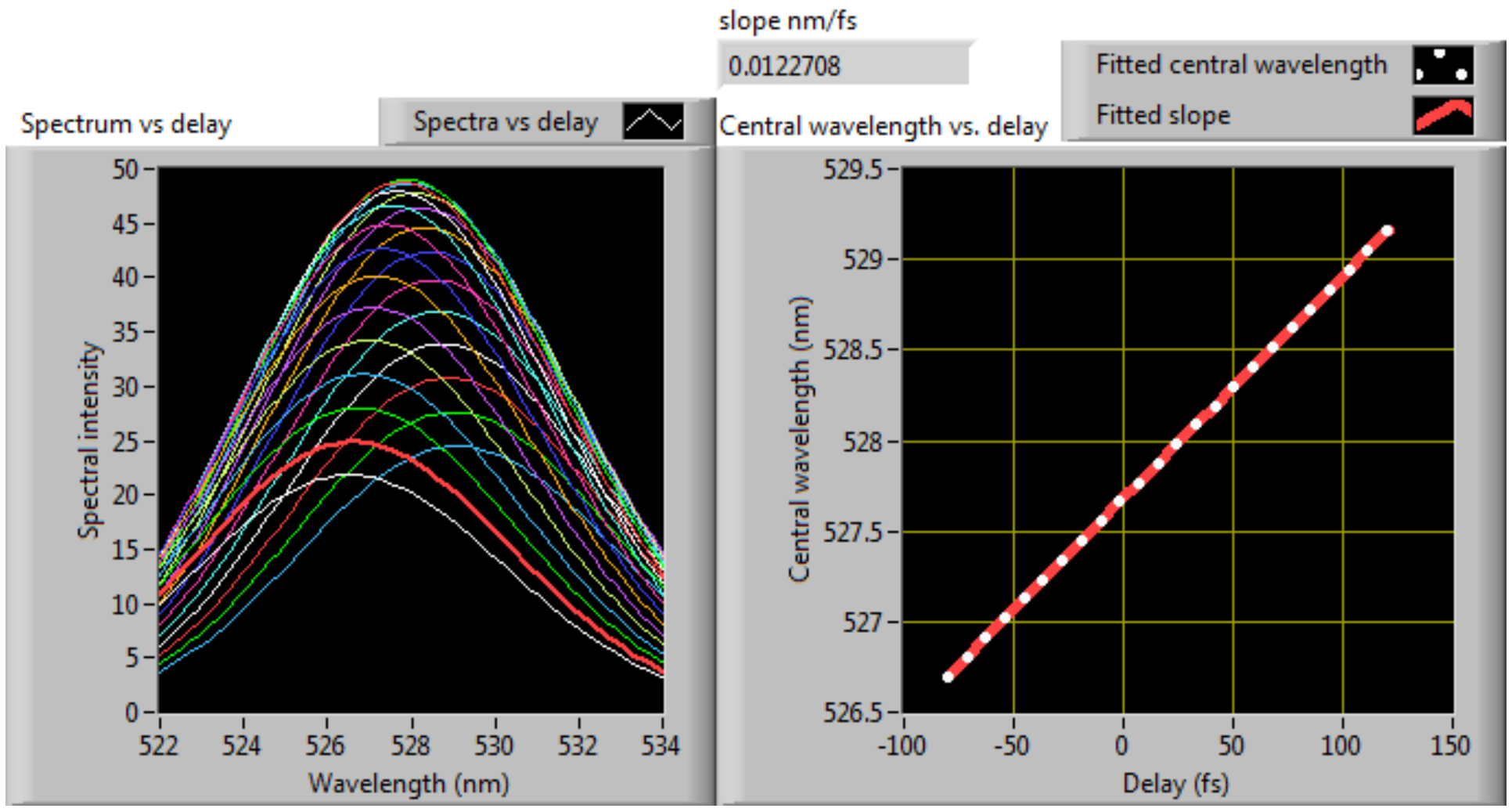

Fig. 9. LabII simulations with the spectrum versus delay between the pulses appearing on the camera shown on the left and the central wavelength of this spectrum with polynomial fit on the right

Figure 10 shows the schematic layout of the integration into the FEL architecture. The reference master clock, operating at 1550nm delivers 180 fs transform limited pulses and is distributed to the clients. The seed oscillator (Ti:Sa) will be optically locked to the reference laser. The electrons will be detected by a beam arrival monitor after the undulators to indicate arrival-time for the X-rays. This signal can also be used to stabilize the accelerator timing. The pump laser timing, compared to the maser clock, will be also corrected after compression to ensure that the achieved relative timing between the X-ray pump and the optical probe is already at its optimum. This is crucial, as several 10's of meters of transfer lines are needed to propagate the laser to the experiment. An X-ray timing tool will then provide accurate and non-invasive measurement of the real timing between the two pulses for binning of the data. As all sub-systems are individually stabilized, the operating range of the timing tool can be reduced, providing higher resolution.

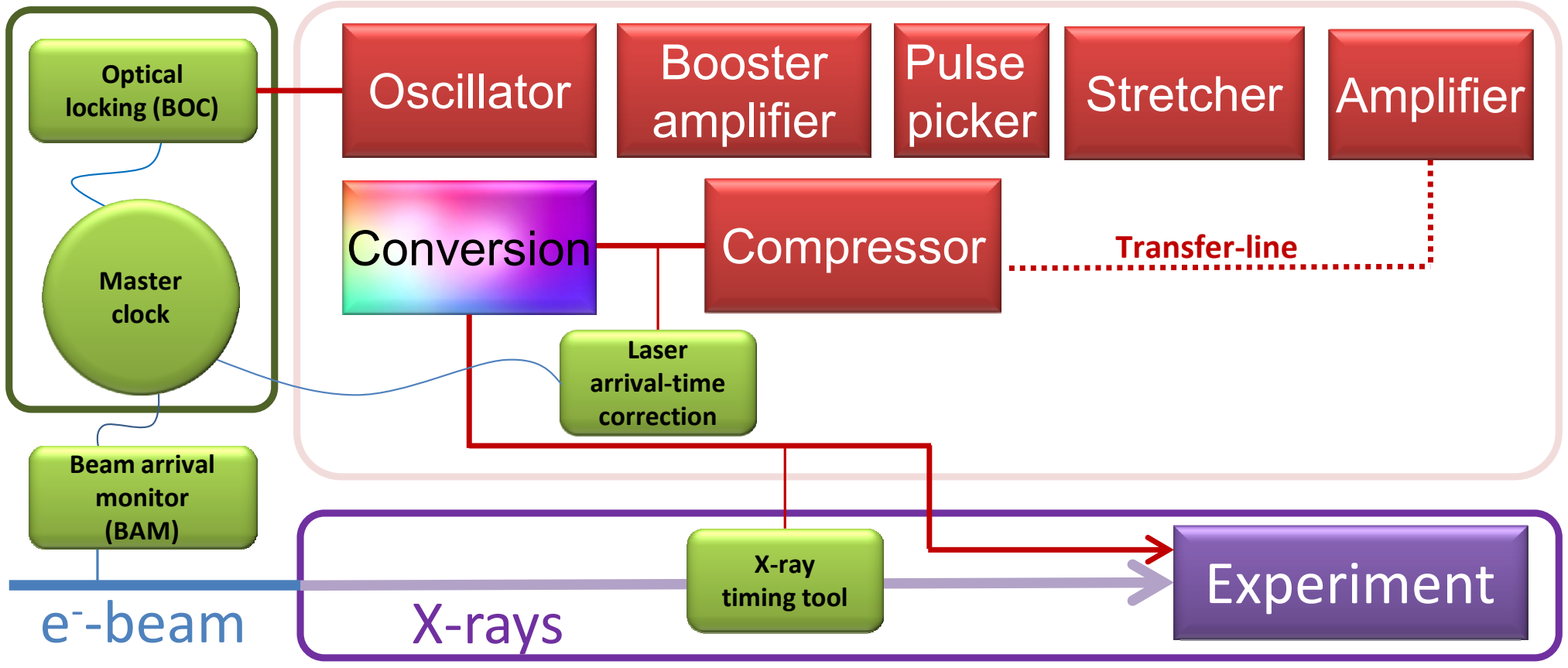

Fig. 10. Schematic of the integration into the FEL experiment. Green areas show the timing related systems, purple the X-ray beam line while read the components of the optical laser system.

## 5. Conclusions

We have developed a laser timing diagnostic and correction system, which can be remotely aligned. The spectrally resolved auto-correlation measurements performed at 800 nm show that with this scheme a resolution of 0.3 fs over the range of the correction (50 fs). Adapting the setup to compare the output of the compressed pulses with the seed oscillator using a non-collinear cross-correlator has given 5 fs resolution. The generated error signal was used to feed back to a delay stage to correct for timing changes in the whole laser chain. The output of the 24 mJ laser system was kept at ~3 fs RMS and 10fs peak to peak jitter compared to the seed oscillator over 10 hours, demonstrating reliable long term operation. The device currently runs at 5 Hz, limited by the communication network. The camera, used in the setup is able to take measurements at the full repetition rate of the SwissFEL (100 Hz) and by analyzing the images locally the jitter can be analyzed at 100 Hz and corrected at Nyquist-frequency. We investigated how to adapt the spectrally resolved setup to mix the output signal directly with the reference clock oscillator of the FEL machine, operating at 1550nm. This system will allow fs level timing correction of the Ti:sapphire probe laser to the FEL experiments, to enable increased resolution for the pump-probe experiments.

**Acknowledgements**

The authors would like to thank the laser team (A. Trisorio and C. Vicario) for their support with the Ti:sapphire system and H. Brands for the camera interface development. We would also like to thank Thomas Feurer for the advice on LabII.